\documentclass[12pt]{article}
\usepackage[latin1]{inputenc}
\usepackage{pstricks}
\usepackage{amsmath}
\usepackage{float}
\usepackage{color}
\input epsf
\usepackage{amssymb}
\usepackage[dvips]{graphicx,psfrag}
\newcommand{\be}{\begin{equation}}
\newcommand{\ee}{\end{equation}}
\newcommand{\bea}{\begin{eqnarray}}
\newcommand{\eea}{\end{eqnarray}}
\newcommand{\lb}{\label}
\begin{document}
\begin{titlepage}
\title{When is the growth index constant?}
\author{
David~Polarski$^1$\thanks{email:david.polarski@umontpellier.fr},
Alexei~A.~Starobinsky$^{2,3}$ \thanks{email:alstar@landau.ac.ru},
Hector~Giacomini$^4$\thanks{email:hector.giacomini@lmpt.univ-tours.fr}
\hfill\\
$^1$ Laboratoire Charles Coulomb, Universit\'e Montpellier 2 \& CNRS\\
 UMR 5221, F-34095 Montpellier, France\\
$^2$ Landau Institute for Theoretical Physics RAS, Moscow, 119334, Russia\\
$^3$ Bogolyubov Laboratory of Theoretical Physics,\\ 
Joint Institute for Nuclear Research, Dubna, 141980, Russia\\
$^4$Universit\'e de Tours, Laboratoire de Math\'ematiques et Physique Th\'eorique,\\
CNRS/UMR 7350, 37200 Tours, France}
\pagestyle{plain}
\date{\today}

\maketitle

\begin{abstract}
The growth index $\gamma$ is an interesting tool to assess the phenomenology 
of dark energy (DE) models, in particular of those beyond general relativity 
(GR). We investigate the possibility for DE models to allow for a constant 
$\gamma$ during the entire matter and DE dominated stages. 
It is shown that if DE is described by quintessence (a scalar field
minimally coupled to gravity), this behaviour of $\gamma$ is excluded
either because it would require a transition to a phantom behaviour 
at some finite moment of time, or, in the case of tracking DE at the 
matter dominated stage, because the relative matter density $\Omega_m$ 
appears to be too small. An infinite number of solutions, with $\Omega_m$ 
and $\gamma$ both constant, are found with $w_{DE}=0$ corresponding to 
Einstein-de Sitter universes.
For all modified gravity DE models satisfying $G_{\rm eff}\ge G$, among them 
the $f(R)$ DE models suggested in the literature, the condition to have a 
constant $w_{DE}$ is strongly violated at the present epoch. 
In contrast, DE tracking dust-like matter deep in the matter era, but with 
$\Omega_m <1$, requires $G_{\rm eff} > G$ and an example is given using 
scalar-tensor gravity for a range of admissible values of $\gamma$. 
For constant $w_{DE}$ inside GR, departure from a quasi-constant value is 
limited until today. Even a large variation of $w_{DE}$ may not result in a 
clear signature in the change of $\gamma$. The change however is substantial 
in the future and the asymptotic value of $\gamma$ is found while its slope 
with respect to $\Omega_m$ (and with respect to $z$) diverges and tends to 
$-\infty$. 
\end{abstract}

\end{titlepage}

\section{Introduction}
The present accelerated expansion of the universe remains a theoretical challenge. A wealth 
of theoretical models and mechanisms were put forward in order to explain it, see the 
reviews \cite{SS00}. 
Remarkably, the simplest model based on GR with a cosmological constant 
$\Lambda$ provides a very good fit to all existing observational data, especially on large 
cosmic scales. Hence this model, apart from the unsolved problem of theoretical derivation 
of $\Lambda$ from quantum field theory, provides a benchmark for the assessment of other 
dark energy (DE) models. One way to make progress is to explore carefully the phenomenology 
of the proposed models and to compare it with observations \cite{WMEHRR13}. It is important 
then to find tools which can efficiently discriminate between models, or between classes of 
models (e.g. \cite{SSS14}). The growth index $\gamma$, which gives a way to parametrize the 
growth of density perturbations in non-relativistic matter component (cold dark matter and 
baryons), is an example of such phenomenological tool. 
This approach was pioneered long time ago in order to discriminate spatially open from 
spatially flat universes \cite{P84} and then generalized to other cases \cite{LLPR91}. It 
was later revived in the context of dark energy \cite{LC07}, with the additional promise to 
single out models formulated outside GR. 
A crucial property is that the growth index has a clear signature in the presence of 
$\Lambda$: the growth index at very low redshifts lies around $0.55$ and it is quasi-constant. 
This behaviour can be extended to smooth noninteracting DE models inside GR with a constant 
equation of state $w_{DE}$, while a strictly constant $\gamma$ is very peculiar \cite{PG07}. 
Such behaviour is strongly violated in some models beyond GR, see e.g. \cite{GMP08,MSY10}. 
In order to gain more understanding, it is interesting to investigate mathematically for which 
models of DE the growth index can be exactly constant, whether inside or outside GR, to see 
if such models are physical and distinguished in some way. That is why this inverse dynamical 
problem is solved below.     
\section{The growth index}
Let us consider the evolution of linear scalar (density) perturbations 
$\delta_m =\delta\rho_m/\rho_m$  in the dust-like matter component 
in the Universe. Deep inside the Hubble radius their dynamics is given by the equation 
\be
{\ddot \delta_m} + 2H {\dot \delta_m} - 4\pi G\rho_m \delta_m = 0~,\label{del}
\ee
where the Hubble parameter $H(t)\equiv \dot a(t)/a(t)$ and $a(t)$ is the scale factor of a 
Friedmann-Lema\i\^ tre-Robertson-Walker (FLRW) universe filled by standard dust-like matter 
and DE components (we neglect
radiation at the matter and DE dominated stages). In the absence of spatial curvature, the 
evolution of the Hubble parameter as a function of the redshift $z=\frac{a_0}{a}-1$ at 
$z\ll z_{eq}$ reads 
\be
h^2(z) = \Omega_{m,0} ~(1+z)^3 + (1 - \Omega_{m,0})
                      ~\exp \left[ 3\int_{0}^z dz'~\frac{1+w_{DE}(z')}{1+z'}\right]~,\lb{h2z}
\ee 
with $h(z)\equiv \frac{H}{H_0}$ and $w_{DE}(z)\equiv p_{DE}(z)/\rho_{DE}(z)$. Equality 
\eqref{h2z} will hold for all FLRW models inside GR. Taking into account that the relative 
density of matter component in terms of the critical one 
$\Omega_m=\Omega_{m,0} \left( \frac{a^3}{a_0^3} h^2 \right)^{-1}$, 
the useful relation follows:
\be 
w_{DE} = - \frac{1}{3(1-\Omega_m)}\, \frac{d\ln \Omega_m}{d\ln (1+z)}\, .  \label{wDE}
\ee
Instead of working with the quantity $\delta_m$, it may be convenient to introduce the 
growth function  $f\equiv \frac{d \ln \delta_m}{d \ln a}$. Then using \eqref{wDE}, it is 
straightforward to show that the equation \eqref{del} leads to the following nonlinear first 
order equation \cite{WS98}
\be
\frac{df}{dx} + f^2 + \frac{1}{2} \left(1 - \frac{d \ln \Omega_m}{dx} \right) f = 
                              \frac{3}{2}~\Omega_m~,\lb{df}
\ee
with $x\equiv \ln a$. The quantity $\delta_m$ is easily recovered from $f$, viz.  
\be
\delta_m(a) = \delta_{m,i}~\exp \left[ \int_{x_i}^{x} f(x') dx' \right]~.
\ee
Clearly $f=p$ if $\delta_m\propto a^p$ (with $p$ constant). In particular $f\to 1$ in 
$\Lambda$CDM for large $z$ and $f=1$ in the Einstein-de Sitter universe. 

In order to characterize the growth of perturbations, the following parametrization 
has been intensively used and investigated in the context of dark energy
\be
f = \Omega_m^{\gamma}~,\lb{Omgamma}
\ee
where $\gamma$ is the growth index. The characterization of the growth of 
matter perturbations using a parametrization of the form (\ref{Omgamma}) has 
attracted a lot of interest with the aim to discriminate between DE models 
based on modified gravity theories and the $\Lambda$CDM paradigm.

We will keep this definition in the general case whenever the growth of matter 
perturbations has no explicit scale dependence and when neither $\Omega_m$, nor 
$\gamma$ are constants: 
\be
f=\Omega_m(z)^{\gamma(z)}~. \lb{gammagen} 
\ee
Maybe rather unexpectedly since no ``conservation law'' for $\gamma$ exists, it appears 
that the growth index is quasi-constant for the standard $\Lambda$CDM, and we will return 
to this point later. Such a behaviour holds for smooth non-interacting DE models when 
$w_{DE}$  is constant, too \cite{PG07}. That is why it is important to investigate whether, 
and when, the growth index $\gamma$ can be exactly constant. We will address this question 
and review some of the results already obtained. An additional interesting point is to 
investigate 
whether and how these results are affected when the evolution of matter perturbations is 
modified. It is known that the behaviour of $\gamma(z)$ can substantially 
differ from its behaviour in $\Lambda$CDM in some modified gravity DE models. 

In many DE models outside GR the modified evolution of matter perturbations has the 
following form
\be
{\ddot \delta_m} + 2H {\dot \delta_m} - 4\pi G_{\rm eff}\rho_m\delta_m = 0~,\label{delmod}
\ee
where $G_{\rm eff}$ is some effective gravitational coupling appearing in the model.
For example, for effectively massless scalar-tensor models \cite{BEPS00}, $G_{\rm eff}$ is 
varying with time but it has no scale dependence while its value today is equal to 
the usual Newton's constant $G$. Introducing for convenience the quantity 
\be
g\equiv \frac{G_{\rm eff}}{G}\lb{g}~,
\ee  
eq.\eqref{delmod} is straightforwardly recast into the modified version of Eq. \eqref{df}, 
viz.
\be
\frac{df}{dx} + f^2 + \frac{1}{2} \left(1 - \frac{d \ln \Omega_m}{dx} \right) f = 
                              \frac{3}{2}~g~\Omega_m~.\lb{dfmod}
\ee
Note that  in \eqref{dfmod}, we keep the same GR definition 
$\Omega_m = \frac{8\pi G \rho_m}{3 H^2}$ in models beyond GR,
i.e. using the Newton gravitational constant $G$, and not $G_{\rm eff}$. As as result, 
$\Omega_m$ defined in this way may exceed unity and then $\rho_{DE}$ becomes negative. 
Then, from \eqref{dfmod} it is straightforward to deduce the following equality 
\be
w_{DE} = - \frac{1}{3(2\gamma -1)} ~\frac{2 \frac{d\gamma}{dx}~\ln \Omega_m  + 
                       1 + 2\Omega_m^{\gamma} - 3 g \Omega_m^{1-\gamma}}{1-\Omega_m}~. \lb{wgen}
\ee
The case $g=1$ reduces to GR.  
Below for any quantity $v$, $v_{\infty}$, resp. $v_{-\infty}$, will denote its (limiting) 
value for $x\to \infty$ in the DE dominated era ($\Omega_m\to 0$), resp. 
$x\to -\infty$ ($\Omega_m\to 1$ unless we have early dark energy, see below). 
In the sequel we will consider models with a constant growth index $\gamma$.
Note that in the opposite case, the first term of the numerator in \eqref{wgen} 
whose magnitude is a priori unknown could become significant. In order to remain finite, 
$\frac{d\gamma}{dx}\big|_{\infty}=-(1+z)~\frac{d\gamma}{dz}\big|_{z=-1} \to 0$ deep in the 
DE domination ($\Omega_{DE}\to 1$).     
\section{A constant growth index inside GR}
Important results can be derived assuming $\gamma$ is constant. We start our analysis 
with models inside GR for which one has 
\bea
w_{DE} &=& - \frac{1}{3(2\gamma -1)} ~\frac{ 1 + 2\Omega_m^{\gamma} - 3 \Omega_m^{1-\gamma}}
                                              {1-\Omega_m} \lb{wGR}\\
   &\equiv&  - \frac{1}{3(2\gamma -1)}~F(\Omega_m; \gamma)~.\lb{F}
\eea  
We see immediately from \eqref{wGR} that $w_{DE}=$ constant is incompatible 
with a constant growth index $\gamma$ inside GR. 

The function $F(\Omega_m;\gamma)$ introduced in \eqref{F} encodes the evolution of 
$w_{DE}$ as a function of $\Omega_m$ from the asymptotic future with 
\be
F(0; \gamma) = 1~, \lb{Ffuture}
\ee
to the asymptotic past with 
\be
F(1; \gamma) = 3 - 5\gamma~,  \lb{Fpast}
\ee 
in case of non-tracking DE in the past ($w_{-\infty}< 0$). Hence we have in the asymptotic 
future ($\Omega_m\to 0$) 
\be
w_{\infty} = -\frac{1}{3(2\gamma -1)}~. \lb{winf}
\ee
The conditions $w_{\infty}<-\frac13$, and $w_{\infty}$ nonsingular, yield the allowed interval
\be
0.5 < \gamma < 1~. \lb{intfDE}
\ee
Though later on the interval \eqref{intfDE} will be refined due to other physical 
considerations, the interesting interval for viable models will remain inside \eqref{intfDE}. 
We note further that $w_{\infty}\approx -1$ is obtained for $\gamma\approx \frac23$ which 
differs substantially from the value realized in $\Lambda$CDM around the present 
epoch. 

In the asymptotic past ($\Omega_m\to 1$), we obtain 
\be
w_{-\infty} = \frac{5\gamma - 3}{3(2\gamma -1)}~. \lb{w-inf}
\ee
As we require $w_{-\infty}<0$, from \eqref{w-inf} and \eqref{intfDE} we must have 
$F(1; \gamma)>0$, and we obtain the refined bounds 
\be
0.5 < \gamma < 0.6~,\lb{intpDE}
\ee 
substantially reducing the allowed interval \eqref{intfDE}. 
For $\gamma=0.6$, DE behaves like dust in the asymptotic past.
 
\begin{figure}
\begin{centering}
\includegraphics[scale=.9]{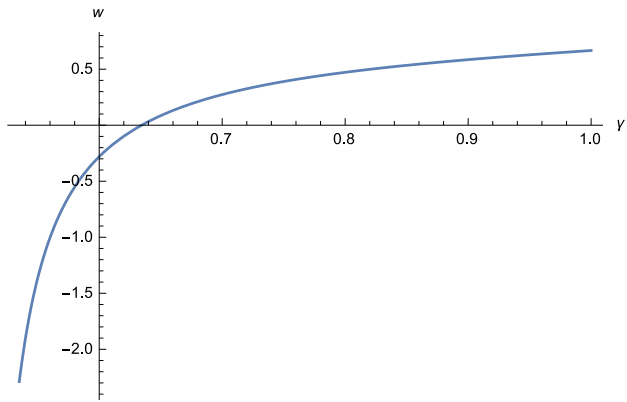}
\par\end{centering}
\caption{The equation of state parameter $w_{DE}$ is displayed as a function of 
$\gamma$ at the present time corresponding to $\Omega_{m,0}=0.30$. We have 
$w_0(0.567628)=-0.8$ and $w_0(0.554094)=-1.2$. Therefore in models 
with a constant growth index $\gamma$, the interval $0.554 \le \gamma \le 0.568$ yields 
a reasonable $w_0$ today. This estimate remains valid when $\gamma$ is 
quasi-constant.}  
\label{fig1}
\end{figure}

We can find a good estimate of the interval yielding a viable EoS at the present 
time. Assuming $-1.2\le w_{DE,0}\equiv w_0\le -0.8$ which covers more than $2\sigma$ 
present observational bounds, we get 
\be
0.554 \le \gamma \le 0.568~, \lb{gam0}
\ee
where we have used $w_0(\gamma=0.554094)=-1.2$ and $w_0(\gamma=0.567628)=-0.8$ and we 
take $\Omega_{m,0}=0.30$, see Figure \ref{fig1}. 
Actually for a constant $\gamma$, all background quantities are given in 
parametric form as follows 
\bea
x(\Omega_m) &=& (2\gamma-1) \int_{\Omega_{m,0}}^{\Omega_m} \frac{d\Omega'_m}
                     {F(\Omega'_m; \gamma) (-\Omega'_m) (1-\Omega'_m)}~, \lb{xOm} \\
t(\Omega_m) &=& (2\gamma-1) H_0^{-1} \int_1^{\Omega_m} \frac{d\Omega'_m}
         {\left[F(\Omega'_m; \gamma) (-\Omega'_m) (1-\Omega'_m)\right]~h(\Omega'_m)}~, \lb{tOm}
\eea
with 
\be
h^2(\Omega_m) = \Omega_{m,0}~\exp \left[-3x(\Omega_m)\right]~\Omega_m^{-1}~.
\ee
For $\gamma$ in the interval \eqref{intpDE}, the following consistent limits 
are indeed obtained 
\bea
x(\Omega_m\to 0) &\sim& (2\gamma-1) \ln \Omega_m^{-1}\to \infty\\
x(\Omega_m\to 1) &\sim& \frac{2\gamma-1}{3-5\gamma} \ln(1-\Omega_m)\to -\infty~.
\eea
Note that integrability of the problem involved is not restricted to the
case of a constant $\gamma$. As was shown in \cite{St98}, the inverse
problem of the determination of the scale factor $a(t)$, the relative Hubble
function $h(z)$ and the corresponding $w_{DE}$ can be solved explicitly
for {\em any} given behaviour of $\delta_m(a)$.

Inside the interval $0.5 < \gamma < \frac{6}{11}$ a phantom behaviour 
is obtained, with $w_{-\infty}\to -\infty$ for $\gamma\to 0.5$. 
Hence for quintessence (a minimally coupled scalar field), for which strong and null 
energy conditions may not be violated, the interval 
$0.5 < \gamma < \frac{6}{11}$ is excluded and the allowed interval for quintessence
reduces to 
\be
\frac{6}{11}\le  \gamma <  0.6~.\lb{int1Q}
\ee 
The value $\gamma = \frac{6}{11}= 0.545454...$ corresponds to $w_{-\infty}=-1$. 
As $w_{\infty}<-1$ for  $0.5 < \gamma < \frac23$, phantom behaviour is unavoidable 
in the future for the interval \eqref{int1Q}. 
\begin{figure}
\begin{centering}
\includegraphics[scale=.85]{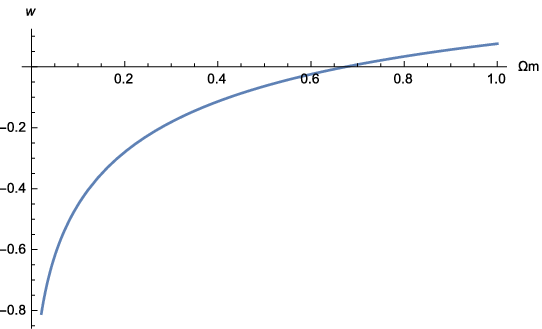}~
\includegraphics[scale=.85]{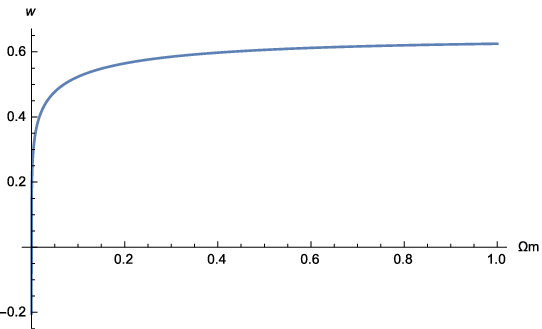}
\par\end{centering}
\caption{The equation of state parameter $w_{DE}$ is displayed as a function of 
$\Omega_m$ for $\gamma=0.61$ (left) and $\gamma=0.9$ (right). Its zero value occurs for 
$\Omega_m<1$ corresponding to a root of $F(\Omega_m; \gamma)$. An infinite number 
of roots for $0.6<\gamma<1$ is obtained. Such roots can be used for a model with tracking DE 
deep in the matter stage. The model displayed on the left is not viable because 
$\Omega_{-\infty}$, the value of $\Omega_m$ corresponding to the root, is too small, 
but a value of $\gamma$ much closer to $0.6$ would produce a viable tracking DE model in 
accordance with \eqref{viable}. The tracking model on the right is not viable because 
$\Omega_{-\infty}$ is vanishingly small, see \eqref{nv}.}  
\label{fig2}
\end{figure}
For the specific case $\gamma=0.6$, we have $w_{-\infty}=0$, however with
$\Omega_ {-\infty}\equiv \Omega_m(x) |_{x\to -\infty} = 1$. Thus, there are no viable non-tracking 
quintessence models for which $\gamma$ is exactly constant \cite{S11}.

But more interesting cases 
do exist in the case of tracking DE in the past when $w_{-\infty}=0$.
Indeed the roots of the equation $F(\Omega_m, \gamma)|_{x\to -\infty}=0$ determine 
the cases when DE has a zero pressure deep inside the matter dominated stage so 
that it is tracking at this stage. 
This can occur for $\Omega_ {-\infty}<1$ if $0.6<\gamma<1$. In particular we have the 
solutions (see Figures \ref{fig2})
\begin{eqnarray}
\gamma &=& \frac{3}{5} \left( 1 +\frac{\varepsilon}{25}\right)~,~~~~~~~~~~~~~~
                                    \varepsilon \equiv 1-\Omega_{-\infty}\ll 1~, \lb{viable} \\
\gamma &=& \frac23~, ~~~~~~~~~~~~~~~~~~~~~~~~~~~~\Omega_ {-\infty}=\frac{1}{8}~, \lb{nvl} \\
\gamma &=& 1-\frac{\ln 3}{\ln \Omega_ {-\infty}^{-1}}~, ~~~~~~~~~~~~~~~ \Omega_{-\infty}\to 0~.
\lb{nv}
\end{eqnarray}
However, while only the upper case \eqref{viable} is consistent with the strong observational 
constraints requiring $\varepsilon < 1\%$ \cite{Pl15}, such universes would 
nevertheless lead to a phantom behaviour in the future, so that quintessence still cannot 
play the role of DE in \eqref{viable}.    
This is avoided for \eqref{nvl}-\eqref{nv}, so these solutions are in principle allowed 
for quintessence.  
Interestingly, for \eqref{nvl} DE tends to a cosmological constant behaviour in the 
future, but this model is not viable because $\Omega_m$ would be too small at the matter 
dominated stage. As $\gamma$ increases, this problem becomes sharper.
As a corollary, universes with a constant $\gamma\ge \frac{2}{3}$ driven by dust and 
quintessence are mathematically possible but excluded by observations.

Finally, the only way that both $\gamma$ and $w_{DE}$ are constant at all times is that 
$\Omega_m$ be always constant as well. This can only be achieved with $w_{DE}=0$ which 
obviously cannot describe our universe dynamics at all times. These universes 
correspond to the roots of $F(\Omega_m; \gamma)$, their expansion is that of 
Einstein-de Sitter universes. Note that these solutions are unstable, the slightest deviation 
of $\Omega_m$, keeping $\gamma$ constant, will lead to a varying $w_{DE}$.   
To summarize, a constant growth index $\gamma$ during the entire evolution of our 
universe is ruled out for constant $w_{DE}$ and for quintessence.  

\section{Solving for the evolution of $\gamma$ inside GR}
While we are mainly interested in a constant $\gamma$ and generically $\gamma$ cannot be 
strictly constant forever, it is interesting that many models actually produce a 
quasi-constant $\gamma$ until today. 
We now turn to the evolution of $\gamma$ for some of these models inside GR. 
Let us consider first the equation governing its dynamics. 
The simplest way is to write it using the variable $\Omega_m$. Then we obtain
\be
2 \alpha ~\Omega_m ~\ln\Omega_m  ~\frac{d\gamma}{d\Omega_m} + \alpha ( 2\gamma - 1 ) 
                         + F(\Omega_m; \gamma) = 0~,\lb{evOm}
\ee
where we have set 
\be
\alpha\equiv 3 ~w_{DE}~. \lb{alpha}
\ee
The entire possible evolution of the universe lies in the interval $\Omega_m=[0,1]$. 

Let us start with  models having a constant $w_{DE}<0$. In this case a constant 
$\gamma$ is obtained numerically in the past starting actually from some low redshifts 
$z\sim 3$ on. Hence in the past we are in the regime where $\gamma$ is constant and we 
can use \eqref{w-inf} in order to relate the initial value 
$\gamma_{-\infty}\equiv \gamma(\Omega_m=1)$ with the constant factor $\alpha < -1$. We 
obtain straightforwardly 
\be
\gamma_{-\infty} = \frac{3 - \alpha}{5 - 2\alpha} = \frac{3(1 - w_{DE})}{5 - 6~w_{DE}}~.\lb{gamma1}
\ee
Hence the natural thing to do from a mathematical point of view is to solve the exact 
equation \eqref{evOm} with the initial condition $\gamma_{-\infty}$ taken at $\Omega_m=1$ (the 
asymptotic past). The solution shows a limited departure from $\gamma_{-\infty}$ around the 
present-day value $\Omega_{m,0}\approx 0.30$. The Taylor expansion around $\Omega_m=1$ up to 
second order is given by  
\bea
\gamma(\Omega_m) &=& \gamma_{-\infty} + \frac{(3-\alpha)(2-\alpha)}{2(2\alpha-5)^2(5-4\alpha)}~
                                                      (1-\Omega_m)\nonumber \\  
     &+& \frac{(3-\alpha)(2-\alpha)(36 \alpha^2 -140 \alpha + 97)}
                 {12(5-2\alpha)^3(5-4\alpha)(5-6\alpha)} ~(1-\Omega_m)^2 + 
                            {\cal O} \left( (1-\Omega_m)^3 \right). \lb{TaylorOm}
\eea
The above expression is remarkably accurate. Actually its accuracy is below the percent level 
already at first order for the present value $\gamma_0=\gamma(\Omega_{m,0})$. 
This means that we happen to live at the epoch where $\gamma$ starts to deviate from 
a constant behaviour. In order to use the quantity $\gamma$ to constrain cosmological models, 
it will be more convenient to use the redshift $z$, or some other variable closely related to 
it, instead of $\Omega_m$. We will return to this point later. Before doing that we will find 
the value of $\gamma$ in the asymptotic future ($\Omega_m=0$). 
This is a mathematically interesting issue. Solving equation \eqref{evOm} in the 
regime $\Omega_m\to 0$, namely 
\be
2 \alpha ~\Omega_m ~\ln \Omega_m ~\frac{d\gamma}{d\Omega_m} + \alpha (2 \gamma - 1) + 1 = 0~, 
\lb{evOmto0}
\ee
the following leading order solution is obtained in the future
\be
\gamma \sim \frac{\alpha-1}{2\alpha} + \frac{C}{\ln \Omega_m}~,~~~~~~~~~~~~~~~~
                                            \Omega_m\to 0, \lb{gamOmto0}
\ee  
where $C$ is some constant. Hence, $\gamma$ tends to the constant value 
\be
\gamma_{\infty}=\frac{\alpha-1}{2\alpha}~, \lb{gaminf}
\ee
which corresponds to the particular constant solution of the inhomegeneous equation 
\eqref{evOmto0}. Obviously, for a non constant $w_{DE}$ with $w_{DE}\to w_{\infty}$, we 
simply substitute $\alpha_{\infty}= 3 w_{\infty}$ in \eqref{gaminf}. 

When this value is substituted in \eqref{winf}, $w_{\infty}$ corresponding to a constant 
$\gamma$ is recovered. 
Hence for a given value $w_{\infty}$, the same asymptotic value $\gamma_{\infty}$ is obtained 
in the future whether we assume $\gamma$ to be constant or not, viz.
\be
\gamma_{\infty}(w) = \gamma(w_{\infty}=w)~, \lb{gg}
\ee
where the left hand side corresponds to a constant $w$ and a varying $\gamma$ while the 
right hand side corresponds to the opposite case. 
\begin{figure}
\begin{centering}
\includegraphics[scale=.85]{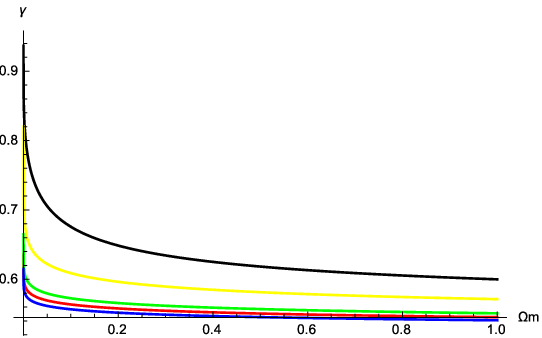}~
\includegraphics[scale=.85]{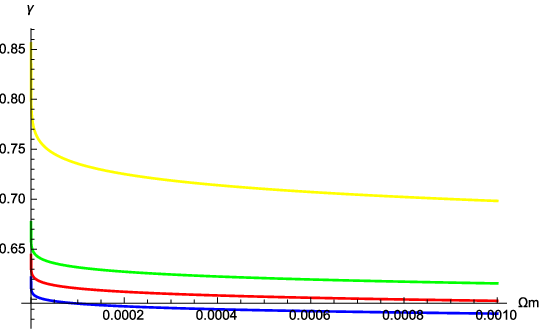}
\par\end{centering}
\caption{The following curves for constant $w_{DE}$ are displayed from bottom to top: 
$w_{DE}=-1.2$ with $\gamma_0\equiv \gamma(\Omega_m=0.30)=0.549062$ (blue); $w_{DE}=-1$ with 
$\gamma_0=0.554727$ (red);
$w_{DE}=-0.8$ with $\gamma_0=0.561896$ (green); $w_{DE}=-\frac{1}{3}$ with $\gamma_0=0.589571$ 
(yellow); $w_{DE}=-\frac{1}{3}\times 10^{-6}$ with $\gamma_0=0.634403$ (black).
In the first four cases it is seen that while a finite value is obtained in the asymptotic 
future, this value is reached with a diverging slope in accordance with equations 
\eqref{gamOmto0} and \eqref{dgamOmto0}. For the black curve with $w_{DE}\approx 0$, a very 
large value is obtained, the value itself would diverge for $w_{DE}\to 0$. We zoom on the 
asymptotic behaviour on the right panel.}
\label{fig3}
\end{figure}
The slope $\frac{d\gamma}{d\Omega_m}$ however diverges with 
\be
\frac{d\gamma}{d\Omega_m}\sim -C \left([\ln \Omega_m]^2~\Omega_m\right)^{-1}~. \lb{dgamOmto0}
\ee 
This behaviour is illustrated in Figures \ref{fig3}.
The derivative with respect to $z$ diverges also.
Note that \eqref{dgamOmto0} implies that $\frac{d\gamma}{d\ln a} = 
\frac{d\gamma}{d\Omega_m} \alpha \Omega_m (1-\Omega_m)\to 0$ for $a\to \infty$, as it was 
emphasized at the end of Section 2 for the quantity $w_{DE}$ to remain finite.  

The same procedure can be applied to the asymptotic past $\Omega_m\to 1$. We write 
\eqref{evOm} in this limit and the solution is then
\be
\gamma \sim \gamma_{-\infty} + \frac{D}{1-\Omega_m}~,
\ee
where $D$ is some integration constant. To avoid a singular behaviour, $D$ must be set to 
zero. Here too, $\gamma_{-\infty}$ corresponds to the particular constant solution of 
\eqref{evOm} in the limit $\Omega_m\to 1$.  

From a practical point of view the use of the redshift $z$ rather than $\Omega_m$ is more 
useful. It is straightforward to rewrite the equation for the evolution of $\gamma$ in terms 
of the variable $y\equiv \frac{a}{a_0}$ which yields
\be
2 \ln \Omega_m~y~\frac{d\gamma}{dy} + \alpha (2\gamma-1)(1-\Omega_m) 
                              + (1-\Omega_m)~F(\Omega_m; \gamma) = 0~. \lb{eva}
\ee
Eq.\eqref{eva} can be solved for models where $\Omega_m$ is known in function of $y$ and 
$w_{DE}$ is dynamical, provided its dynamics in terms of $y$ is known as 
well. Actually, if $w_{DE}(y)$ is known, then the background evolution, and hence 
$\Omega_m(y)$, is completely specified. 

Let us start with a constant $w_{DE}=w_0$. We can follow the same procedure as before, 
taking the initial condition \eqref{gamma1}. Of primary interest for the confrontation with 
observations is the evolution of $\gamma$ around the present time, i.e. around $y=1$. 
Written in terms of $z$ this expansion becomes
\be
\gamma = \gamma_0 + \gamma^{(1)}~\frac{z}{1+z} +  
                    {\cal O} \left[ \left(\frac{z}{1+z}\right)^2 \right]~, \lb{Taylorz}
\ee
with
\be
\gamma^{(1)} = \frac{1-\Omega_{m,0}}{2\ln\Omega_{m,0}}~\Big[3 w_0 (2\gamma_0-1) + 
              F(\Omega_{m,0};\gamma_0)\Big]~.  \lb{gam1}
\ee
The present-day value $\gamma_0$ must be found \emph{numerically} using Eq.\eqref{eva}.   
Like for the expansion \eqref{TaylorOm}, \eqref{Taylorz} has an accuracy below the percent 
level already at first order up to $z\approx 3$. 
We stress that $\gamma^{(1)}$ is a known function of the underlying parameters $\Omega_{m,0}$ 
and $w_0$, not an additional free parameter, and in particular that it depends also on 
$\gamma_0$.    

The useful expansion is really in terms of the variable $1-y\equiv \frac{z}{1+z}$. 
The advantages of this variable is well-known in the cosmographic approach, and this 
variable is actually used for the definition of the (CPL) parametrization \cite{CP01}
\be
w_{DE}(y) = -1+ A + B(1-y)\equiv w_0 + w_a (1-y)~.\lb{par}
\ee 
Next step is to consider dynamical dark energy models. Eq.\eqref{eva} is readily 
applied to any parametrized model $w_{DE}(y)$. 
The motivation here is to use a parametrised expression for the underlying equation of 
state $w_{DE}$, and then to use the accurate result \eqref{Taylorz} instead of looking 
for some parametrized form for $\gamma$.

It is straightforward to find the first two terms of the expansion \eqref{Taylorz} for 
the CPL model. The value $\gamma_0$ has to be found numerically and will depend on both 
parameters. We can use the same procedure as for a constant EoS in order to find the initial 
value $\gamma_{-\infty}$ because in the past $w_{CPL}$ tends to a constant value so that we get 
now
\be
\gamma_{-\infty} = \frac{6 - 3(A+B)}{11 - 6(A+B)}= \frac{3[1-(w_0+w_a)]}{5-6(w_0+w_a)}~.
\lb{gamma1par}
\ee
The quantities $\gamma_0$ and $\gamma_{-\infty}$ depend on both parameters $w_0$ and $w_a$. 
In contrast, the coefficient $\gamma^{(1)}$ is again given by \eqref{gam1} and does 
not depend explicitly on $B$ ($w_a$).
Indeed the coefficient $B$ does not appear explicitly at first order of the expansion 
\eqref{Taylorz} for the parametrization \eqref{par}. The expression \eqref{Taylorz} 
here too is very accurate already at first order up to $z\sim 3$.
For example, for $A=B=0.1$ and $\Omega_{m,0}=0.30$, we obtain numerically 
$\gamma(z=3)=0.550922$. 
If we use the expansion \eqref{Taylorz} with \eqref{gam1}, one obtains 
$\gamma(z=3)=0.548222$ with $\gamma_0=0.558862$ obtained numerically. 
One should compare these numbers to the value $\gamma_{-\infty}=0.551020$ and we note that 
$\gamma(z=3)$ is already very close to this value. 
A variation of the EoS will hardly be observable for a modest change in $\gamma$.

\section{A constant growth index beyond GR}
As modified gravity DE models are known to allow for an effective DE component of 
the phantom type \cite{BEPS00,MG}, they cannot be ruled out for the same reason as 
quintessence. 
One could expect that these models offer better prospects to accommodate a constant $\gamma$, 
possibly with a constant $w_{DE}$ or with tracking DE and we turn now our attention to these 
questions.

For constant $\gamma$ and $g\ne 1$ our starting point is
\be
w_{DE} = - \frac{1}{3(2\gamma -1)} ~\frac{ 1 + 2\Omega_m^{\gamma} - 3 g \Omega_m^{1-\gamma}}
                                              {1-\Omega_m}~. \lb{wmod}
\ee
We see that $w_{\infty}$ is again given by \eqref{winf} with $\gamma$ in the interval 
\eqref{intfDE}. 
Looking at the asymptotic past, we obtain 
\be
w_{-\infty} = \frac{5\gamma - 3}{3(2\gamma -1)} + b~, \lb{wmod-inf}
\ee
where we have defined 
\be
b\equiv - \frac{1}{2\gamma -1}~\left[\frac{dg}{d\Omega_m}\right]_{-\infty}~. \lb{b}
\ee
Clearly $g_{-\infty}=1$ to avoid that $w_{-\infty}$ diverges. We have therefore 
$\left[\frac{d^ng}{dx^n}\right]_{-\infty}=0~\forall n$. 
The condition $w_{-\infty}< 0$ together with \eqref{intfDE} yields  
\be
0.5 < \gamma < \gamma_{\rm max}(b)~, \lb{int1DE_b}
\ee
with $\gamma_{\rm max}(b)\equiv \frac{ 3b+3 }{ 6b+5 },~ b>-\frac23$.
For $b<0$, we have $0.6<\gamma_{\rm max}(b)<1$ for $-\frac{2}{3}<b<0$.
As expected, the allowed interval \eqref{intpDE} is re-obtained for $b=0$ and 
for $b>0$ \eqref{int1DE_b} shrinks to zero when $b\to \infty$. 
Hence the interval \eqref{intfDE} is not necessarily reduced by 
constraints from the past universe. 
Further $w_{\infty}$ is no longer of the phantom type 
if $\frac{2}{3}\le \gamma <1$, which is possible for 
$-\frac{2}{3} < b \le -\frac{1}{3}$.

\subsection{A constant $w_{DE}$}
We derive now the condition for $w_{DE}$ to be constant.
It is easy to see that this is achieved provided the following equality hold
\be
\Omega_m + 2 \Omega_m^{\gamma} - 3 g \Omega_m^{1-\gamma} = 0~, \lb{condg} 
\ee
which constrains the evolution of $g$ as follows  
\be
g = \frac13 \left(\Omega_m^{\gamma} + 2 \Omega_m^{2\gamma - 1} \right)~. \lb{gOm}
\ee
As a result of \eqref{gOm} we have $g \to 0$ for $\Omega_m\to 0$ and again $g\to 1$ 
when $\Omega_m\to 1$. A powerful constraint on $g$ is implied by \eqref{gOm}, viz. 
\be
g < 1~,~~~~~~~~~~~~~~~{\rm for}~~\Omega_m<1~. \lb{gl1}
\ee 
In the DE domination it is clear that \eqref{gOm} would be strongly 
violated for $g=1$ and this violation will be even stronger for $g>1$.  
According to Eq.\eqref{gOm}, $g$ would eventually vanish in a DE dominated universe. We 
recover in particular the result that a constant growth index is not possible inside GR for 
a constant equation of state $w_{DE}$. Interestingly, a value $\gamma_0 <0.5$ would also 
increase the r.h.s. of \eqref{gOm} at the present epoch, but $\gamma$ cannot have this value 
during the entire evolution of the universe.   
 
We can also use \eqref{gOm} in a different perspective: knowing that $w_{DE}$ is constant, is 
it possible to have a constant $\gamma$? We see from \eqref{gOm} that this is possible for 
$g=1$ and $\Omega_m\approx 1$. So a constant $w_{DE}$ is compatible with a constant 
$\gamma$ inside GR deep in the matter era with $w_{DE} = w_{-\infty}$ and the corresponding 
$\gamma$ is found from eq.\eqref{w-inf}. 
It is interesting to relate this result to $\Lambda$CDM. On very low redshifts, 
$z\lesssim 0.5$, we obtain to high accuracy \cite{PG07} 
$\gamma(z)\simeq \gamma_0(\Omega_{m,0}) - 0.02~z$ 
with $\gamma_0\approx 0.55$ depending on $\Omega_{m,0}\approx 0.3$. 
While in the DE dominated stage, $\Lambda$CDM would violate \eqref{gOm} (with $g=1$), 
on larger redshifts $\gamma$ tends to the slightly lower asymptotic value 
$\gamma_{-\infty} = \frac{6}{11}$ which is essentially reached for $z\gtrsim 3$ 
confirming our results. 

It is also interesting to consider a situation where $w_{DE}=w_0$ is constant 
(but $w_0\ne w_{\infty}$) on a restricted part of the expansion only which we take around the 
present time. Then during that stage, the relation \eqref{gOm} taken at the present time is 
generalized to 
\be
g_0 \equiv 1= \frac13 ~\left[ (1-\beta) ~\Omega_{m,0}^{\gamma -1} + \beta~\Omega_{m,0}^{\gamma} + 
                           2 ~\Omega_{m,0}^{2\gamma -1} \right]~,\lb{gOmalpha} 
\ee 
where we have defined $\beta\equiv -3 w_0 (2\gamma -1)$. Eq.\eqref{gOm} taken today is 
recovered for $\beta=1$. Then again \eqref{gOmalpha} would be generically strongly violated 
if a constant $\gamma$ is kept around the present time. 
This is particularly interesting for all modified gravity DE models which are known to produce 
$w_0\approx -1$ at the present epoch.
For example, for $w_0=-1$, $\gamma=0.6$, and $\Omega_{m,0}=0.30$, we obtain $g_0=0.90$, still 
significantly less than one. This is to be compared with $g_0=0.68$ for $\beta=1$. 

We want to comment finally on a possible scale dependence of the growth index.
In many modified gravity DE models \cite{MG} the appearance of a fifth force, and its 
effective screening on small scales where this fifth force should not be felt, implies 
generically a scale dependence of $g$ and therefore of the perturbation growth on large 
cosmic scales. In that case we can still use \eqref{wmod}, however for each scale $k^{-1}$ 
separately and the growth index $\gamma$, though constant in time, will differ for different 
scales. All the results derived above hold for each scale separately. 

Let us now apply our results to some concrete models outside GR. We choose one model, a model 
based on scalar-tensor gravity where the growth of perturbations deep inside the Hubble 
radius is scale-independent and another family of models, $f(R)$ DE models, where the growth 
is scale-dependent, too.
\vskip 10pt
\par\noindent
a) We consider first the scalar-tensor (ST) DE model considered in \cite{BEPS00} with 
Lagrangian 
\be
L=\frac12 \Bigl( F(\Phi)~R - 
Z(\Phi)~g^{\mu\nu}\partial_{\mu}\Phi\partial_{\nu}\Phi \Bigr) - U(\Phi) + L_m(g_{\mu\nu})~,
\ee 
with a version of this model which is essentially massless on cosmic scales
\footnote{One should not confuse $F(\Phi)$ with the function $F(\Omega_m;\gamma)$ introduced 
in Eq. \eqref{F}, nor $f(R)$ with the growth function $f$ introduced in Eq. \eqref{df}.}.
In this unscreened version of the model, the quantity $G_{\rm eff}$ affecting the perturbations 
growth through \eqref{delmod} corresponds also to the gravitational coupling between two 
test masses in a laboratory experiment and it is given by the expression 
$G_{\rm eff}= \frac{1}{8\pi F} \bigl(1 + \frac{1}{2\omega_{BD} + 3} \bigr)$. 
Hence its value today $G_{{\rm eff},0}$ is equal to the Newton constant $G$, and we have today 
$g=1$. By having today a very large Brans-Dicke parameter $\omega_{BD,0}$, these models obey 
laboratory and Solar system constraints \cite{GPRS06} with 
$G=G_{{\rm eff},0}\approx \frac{1}{8\pi F_0}$. For $\Omega_{m,0}\approx 0.30$, Eq.\eqref{gOm} 
is grossly violated today, so this model cannot accommodate a constant growth 
index $\gamma$ with a constant $w_{DE}$. 
\vskip 10pt
\par\noindent
b) An interesting generalization can be given for some modified gravity models like f(R) DE 
models (see e.g. the review \cite{FT10}). These models comply with laboratory and Solar system 
constraints due to screening of the fifth force for scales that are larger than some critical 
scale $\lambda_c(R)$. This scale is the Compton length of the scalaron -- a scalar degree of 
freedom, or particle in quantum language, appearing in $f(R)$ gravity -- which depends on the 
Ricci scalar $R$ and finally on the matter density \footnote{This behaviour is well
known in plasma physics (plasmon) and in elementary particle physics, still in cosmology the 
special term "chameleon" is often introduced.}. 
Then the growth of matter density perturbations obeys 
Eq. \eqref{delmod}, with $G_{\rm eff}(z,k)$ being both time and scale dependent. 
Due to this screening mechanism, $G_{\rm eff}(z,k)$ reduces to $G$ in high curvature regions. 
Let us see explicitly how this works for viable $f(R)$ DE models suggested in the literature. 
We then have \cite{Z06,T07}:
\be
g(z,k) = \left( \frac{df}{dR} \right)^{-1}~\left[ 1 +
\frac{\left(\frac{\lambda_c}{\lambda}\right)^2 }{3\left(1 +
\left(\frac{\lambda_c}{\lambda}\right)^2 \right)} \right],
~~\lambda = \frac{a(t)}{k}~. \lb{gfR}
\ee 
In viable $f(R)$ models of present DE, all relevant cosmic scales  satisfy 
$\lambda \gg \lambda_c(R)$ at the matter era  with $\frac{df}{dR}=1$ to high accuracy. In this 
way the standard growth of perturbations is regained. At low redshifts however, as the 
critical length $\lambda_c$ increases significantly with the decrease of matter density and 
the Ricci scalar $R$, for cosmic scales smaller than $\lambda_c$ we will 
have $G_{\rm eff}(z,k)>G$ and matter perturbations on these scales will experience a modified 
(boosted) growth. As $\frac{d^2f}{dR^2}>0$ \cite{S07,ABS10} (otherwise a weak curvature 
singularity appears in solutions), the factor in front of the brackets in 
\eqref{gfR} increases as the universe expands and it becomes larger than one. So we have for 
these models $g\ge 1$ for any scale at any time. The quantity $g$ becomes as large as $4/3$
in the present era on scales $\lambda<\lambda_c$ where the growth of 
matter perturbations is boosted, so that \eqref{gOm} gets strongly violated. 
Actually, for any modified gravity model with $g\ge 1$ 
and for $w_{DE}\approx -1$ in the DE domination, \eqref{gOm} or \eqref{gOmalpha} is 
strongly violated at the present time. It is interesting in this respect that generalized 
Proca DE models were recently suggested for which $g<1$ seems possible \cite{FHKMTZ16}.  

\subsection{Tracking dark energy in the matter era}

While modified gravity DE models with $g\ge 1$ cannot accommodate a constant $w_{DE}$, 
let us consider if they allow DE to scale like dust-like matter deep in the matter era with 
$\Omega_{m,-\infty} < 1$. From \eqref{wmod}, the following equation should be satisfied  
\be
1 + 2 \Omega_{m,-\infty}^{\gamma} - 3 g_{-\infty} \Omega_{m,-\infty}^{1-\gamma} = 0~.  \lb{condEDE}
\ee
Equation \eqref{condEDE} has solutions only for $g_{-\infty}>1$. It is seen that in this 
case $g_{-\infty}$ is no longer constrained to be equal to one. 
If we take any realistic $\Omega_{m,-\infty}<1$, but close to $1$, and $g_{-\infty}>1$, a 
$\gamma$ can be found which solves \eqref{condEDE}
\be
\gamma = \gamma ( \Omega_{m,-\infty}; g_{-\infty})~. \lb{gammag}
\ee
It is also seen from \eqref{gammag} that $\gamma$ is not unique but depends on the model 
parameters $\Omega_{m,-\infty}$ and $g_{-\infty}$. In case $g$ is scale dependent, \eqref{gammag} 
should be considered for each scale separately. 

We will show now how this can be realized using ST gravity. 
As was shown in \cite{GPRS06}, one can construct a so-called asymptotically safe version for 
which $F$ tends in the past to a constant value $F_{-\infty}$. So we have in the asymptotic 
past 
\be
G_{-\infty} = G_{{\rm eff},-\infty} = \frac{1}{8\pi F_{-\infty}}, \lb{Gas}
\ee
the model tends to GR in the past however with the gravitational constant \eqref{Gas} 
different from its present-day value $\frac{1}{8\pi F_0}$ which is used in the 
definition of the relative densities $\Omega_i\equiv \frac{\rho_i}{3 F_0 H^2}$. 
Matter perturbations on scales deep enough inside the Hubble radius in the past obey 
eq.\eqref{delmod} with a constant $G_{\rm eff}$ given by \eqref{Gas}. 
In this scenario one has deep in the matter era 
\bea
F \to  F_{-\infty} = {\rm const},~~~
|\dot F| \ll HF_{-\infty}~,~~~~|\ddot F|\ll H^2F_{-\infty}~,~~~~H^2 \propto (1+z)^3~.\lb{asst}
\eea
We get from the modified Friedmann equations
\be
\Omega_{DE,-\infty} = 1 - \Omega_{m,-\infty} = 
            2\Omega_{U,-\infty} + \left( 1  - \frac{F_{-\infty}}{F_0}\right)~,\lb{OmDE-infty} 
\ee
where $\Omega_{U,-\infty}\equiv \frac{U_{-\infty}}{3 F_0 H^2}$.
Note that Big-Bang nucleosynthesis bound constraints $F_{-\infty}$ to be close to $F_0$ and we 
take $F_{-\infty}<F_0$. So we have in particular 
\be
g_{-\infty} = \frac{F_0}{F_{-\infty}} > 1~.
\ee
So we see that a constant $\gamma$ \eqref{gammag} can be found which is a solution of 
Eq.\eqref{condEDE}. Hence on those scales for which matter perturbations satisfy the modified 
growth of perturbations with constant $G_{\rm eff}$ given by \eqref{Gas}, a constant 
$\gamma$ is compatible with DE behaving like dust while $\Omega_m < 1$.  

Interestingly, a second family of tracking solutions are allowed in this model. Indeed, if we 
take $F \to  F_{-\infty} > F_0$ (but again not too much larger), so that now 
$g_{-\infty} = \frac{F_0}{F_{-\infty}} < 1$ and $\Omega_{m,-\infty}>1$, \eqref{condEDE} will have 
solutions corresponding to this case too leading to new tracking solutions. 
\section{Conclusion}

In this work we have studied the conditions under which the growth index $\gamma$ of scalar 
(density)  perturbations in the non-relativistic matter component in the Universe (cold dark 
matter and baryons) can be constant in the presence of DE not interacting directly with 
matter \footnote{For complementary approaches to the growth index and its use, 
see e.g. \cite{gamma}.}. 
We emphasize that we mean by that the condition for $\gamma$ to be strictly constant 
during the entire evolution of the Universe after the end of the radiation dominated stage. 
We have also investigated the possibility for DE models to accommodate for both a constant 
$\gamma$ and a constant DE equation of state $w_{DE}$. It appears that if DE is described by 
quintessence (a scalar field minimally coupled to gravity) in the GR framework, this behaviour 
of $\gamma$ is excluded either because it would require transition to phantom 
behaviour of DE at some finite moment of time (which is not possible for quintessence), or, in 
the case of early (tracking) DE at the matter dominated stage, because the relative matter 
density $\Omega_m$ appears to be too small. 
Thus, it seems to be nothing more deep in $\gamma$ being exactly constant, and its 
quasi-constant behaviour in the standard $\Lambda$CDM and the simplest quintessence models of 
DE is simply a consequence of a narrow allowed interval for its variation, Eq. \eqref{int1Q}, 
even with adding the case Eq. \eqref{viable} for early DE.  
We have found that this is also ruled out in many modified gravity models. 
We have shown it explicitly for massless scalar-tensor 
DE models and for the $f(R)$ DE models suggested in the literature. 
In all these models, the condition \eqref{gOm} would be strongly violated at the DE 
domination, and even stronger on cosmic scales where the growth of matter perturbations 
is boosted.
Assuming $w_{DE}=-1$ around the present time only would still lead to a strong violation 
of \eqref{gOm} today. It is interesting that in DE models beyond GR allowing $g<1$, 
like generalized Proca models, violation of \eqref{gOm} today could be milder.    
Of course, $w_{DE}\approx -1$ around the present time is possible in these models if $\gamma$ 
is substantially non constant. We believe that the results obtained here, which relate the 
behaviour of $\gamma$ with the effective gravitational coupling, shed a new light on 
earlier findings for $f(R)$ models \cite{GMP08, MSY10} where the growth index showed 
substantial variation already on very low $z$ with $\gamma_0<0.5$. This is interesting in 
particular as a nonstandard behaviour of the growth index $\gamma$ could be one of the 
smoking guns of DE models beyond GR.

\section*{Acknowledgments}
A.A.S. was partially supported by the grant RFBR 14-02-00894 and by the Scientific Program P-7 
of the Presidium of the Russian Academy of Sciences. D.P. acknowledges useful discussions with 
Stefan Auclair. 
 


\begin{thebibliography}{99}

\bibitem{SS00} V.~Sahni and A.~A.~Starobinsky, Int. J. Mod. Phys. D {\bf 9}, 373 (2000);   
P.~J.~E.~Peebles and B.~Ratra, Rev. Mod. Phys. {\bf 75}, 559 (2003); 
E.~J.~Copeland, M.~Sami and S.~Tsujikawa, Int. J. Mod. Phys. D {\bf 15}, 1753 (2006);
V.~Sahni and A.~A.~Starobinsky, Int. J. Mod. Phys. {\bf 15}, 2105 (2006);
 M.~Li, X.-D.~Li, S.~Wang and Y.~Wang, Commun. Theor. Phys. {\bf 56}, 525 (2011).

\bibitem{WMEHRR13} D.~H.~Weinberg, M.~J.~Mortonson, D.~J.~Eisenstein, C.~Hirata, A.~G.~Riess 
and E.~Rozo, Phys. Rept. {\bf 530}, 87 (2013); 
L. Amendola {\it et al.}, Living Rev. Rel. {\bf16}, 6 (2013); 
P. Bull {\it et al.}, Phys. Dark Univ. {\bf 12}, 56 (2016).

\bibitem{SSS14} V.~Sahni, A.~Shafieloo and A.~A.~Starobinsky, Astrophys. J. {\bf 793}, 
 L40 (2014).

\bibitem{P84} P.~J.~E.~Peebles, Astrophys. J. {\bf 284}, 439 (1984).

\bibitem{LLPR91} O.~Lahav, P.~B.~Lilje, J.~R.~Primack and M.~J.~Rees, MNRAS {\bf 251}, 128 
 (1991).

\bibitem{LC07} E.~V.~Linder and R.~N.~Cahn, Astropart. Phys. {\bf 28} 481 (2007). 

\bibitem{PG07} D. Polarski and R. Gannouji, Phys. Lett. B {\bf 660}, 439 (2008).

\bibitem{GMP08} R.~Gannouji, B.~Moraes and D.~Polarski, JCAP {\bf 0902}, 034 (2009).

\bibitem{MSY10} H.~Motohashi, A.~A.~Starobinsky and J.~Yokoyama, Progr. Theor. 
 Phys. {\bf 123}, 887 (2010).

\bibitem{WS98} L.~Wang and P.~J.~Steinhardt, Astrophys. J. {\bf 508}, 483 (1998).

\bibitem{BEPS00} B.~Boisseau, G.~Esposito-Far\`ese, D.~Polarski and A.~A.~Starobinsky,
 Phys. Rev. Lett. {\bf 85}, 2236 (2000).

\bibitem{St98} A.~A.~Starobinsky, JETP Lett. 68, 757 (1998) [arXiv:astro-ph/9810431].

\bibitem{S11} A.~A.~Starobinsky, invited talks at the conferences Cosmology Workshop 
Montpellier11 (Montpellier, 03.11.2011) and  HEA-2011 (Moscow, 14.12.2011), unpublished.

\bibitem{Pl15} P.~A~R.~Ade {\it et al.}, Astron. Astroph. {\bf 594}, A13 (2016).

\bibitem{CP01} M. Chevallier and D. Polarski, Int. J. Mod. Phys. D{\bf 10}, 213 (2001);
 E. V. Linder, Phys. Rev. Lett. {\bf 90}, 091301 (2003).

\bibitem{MG} T.~Clifton, P.~G.~Ferreira, A.~Padilla and  C.~Skordis, Phys. Rept. {\bf 513}, 
 1 (2012); 
 A.~Joyce, B.~Jain, J.~Khoury and  M.~Trodden, Phys. Rept. {\bf 568}, 1 (2015).

\bibitem{GPRS06} R.~Gannouji, D.~Polarski, A.~Ranquet and A.~A.~Starobinsky, JCAP {\bf 0609}, 
 016 (2006).

\bibitem{FT10} A.~De Felice and S.~Tsujikawa, Living Rev. Rel. {\bf 13}, 3 (2010).

\bibitem{Z06} P.~Zhang, Phys. Rev. D {\bf 73}, 123504 (2006).

\bibitem{T07} S.~Tsujikawa, Phys. Rev. D {\bf 76}, 023514 (2007).

\bibitem{S07} A.~A.~Starobinsky, JETP Lett.{\bf 86}, 157 (2007).

\bibitem{ABS10} S.~A.~Appleby, R.~A.~Battye and A.~A.~Starobinsky, JCAP {\bf 1006}, 005 (2010).

\bibitem{FHKMTZ16} A.~De Felice, L.~Heisenberg, R.~Kase, S.~Mukohyama, S.~Tsujikawa and  
Y.~Zhang, Phys. Rev. D {\bf  94}, 044024 (2016).

\bibitem{gamma} 
M. Malekjani, S. Basilakos, Z. Davari, A. Mehrabi, M. Rezaei, 
Mon. Not. Roy. Astron. Soc. 464, 1192 (2017);\\
Alberto Bailoni, Alessio Spurio Mancini, Luca Amendola, arXiv:1608.00458;\\
Xiao-Wei Duan, Min Zhou, Tong-Jie Zhang, arXiv:1605.03947;\\
B. Wang, E. Abdalla, F. Atrio-Barandela, D. Pavon, Rept. Prog. Phys. 79, no.9, 096901 (2016);\\ 
N. Nazari-Pooya, M. Malekjani, F. Pace, D. Mohammad-Zadeh Jassur, 
Mon. Not. Roy. Astron. Soc. 458, no.4, 3795 (2016);\\ 
S. Basilakos, J. Solà, Phys. Rev. D92, no.12, 123501 (2015);\\
I. de Martino, M. De Laurentis, S. Capozziello, Universe 1, no.2, 123 (2015);\\
J. N. Dossett, M. Ishak, D. Parkinson, T. Davis, Phys.Rev. D92, no.2, 023003 (2015);\\ 
A. B. Mantz et al., Mon. Not. Roy. Astron. Soc. 446, 2205 (2015);\\ 
S. Nesseris, S. Basilakos, E.N. Saridakis, L. Perivolaropoulos, Phys. Rev. D88 103010 (2013);\\ 
K. Bamba, Antonio Lopez-Revelles, R. Myrzakulov, S.D. Odintsov, L. Sebastiani, 
Class. Quant. Grav. 30 015008 (2013);\\ 
A. Bueno belloso, J. Garcia-Bellido, D. Sapone, JCAP 1110, 010 (2011);\\ 
R. Bean, M Tangmatitham, Phys. Rev. D81, 083534 (2010);\\
Puxun Wu, Hong Wei Yu, Xiangyun Fu, JCAP 0906, 019 (2009);\\
Seokcheon Lee, Kin-Wang Ng, Phys. Lett. B688, 1 (2010);\\ 
Yungui Gong, Phys.Rev. D78, 123010 (2008);\\ 
V. Acquaviva, A. Hajian, D. N. Spergel, S. Das, Phys. Rev. D78 043514 (2008);\\ 
Hao Wei, Phys. Lett. B664 1 (2008);\\ 
S. Nesseris, L. Perivolaropoulos, Phys. Rev. D77, 023504 (2008).

\end{thebibliography}
\end{document}